\documentclass[aps,amsmath,superscriptaddress]{revtex4-1}
\usepackage{amsmath}
\usepackage{amssymb}
\usepackage[pdftex,bookmarks,colorlinks]{hyperref}
\pdfoutput=1
\usepackage[pdftex]{graphicx}
\usepackage{blindtext}
\usepackage{natbib}
\usepackage{upgreek}
\usepackage{subcaption}
\usepackage{calc}
\usepackage{amsmath}
\usepackage{amssymb}
\usepackage{array}

\begin{document}
\title{Supressed splashing on elastic membranes}
\author{Marise V. Gielen} \email[]{m.v.gielen@utwente.nl} 
\author{Ri\"elle de Ruiter} 
\affiliation{Physics of Fluids Group, Max Planck Center Twente for Complex Fluid Dynamics, JM Burgers Center, and MESA+ Center for Nanotechnology, Department of Science and Technology, University of Twente, P.O. Box 217, 7500 AE Enschede, The Netherlands.}
\author{Jacco H. Snoeijer}
\affiliation{Physics of Fluids Group, Max Planck Center Twente for Complex Fluid Dynamics, JM Burgers Center, and MESA+ Center for Nanotechnology, Department of Science and Technology, University of Twente, P.O. Box 217, 7500 AE Enschede, The Netherlands.}
\affiliation{Mesoscopic Transport Phenomena, Eindhoven University of Technology, P.O. Box 513, 5300 MB Eindhoven, The Netherlands.}
\author{Hanneke Gelderblom} \email[]{h.gelderblom@utwente.nl}
\affiliation{Physics of Fluids Group, Max Planck Center Twente for Complex Fluid Dynamics, JM Burgers Center, and MESA+ Center for Nanotechnology, Department of Science and Technology, University of Twente, P.O. Box 217, 7500 AE Enschede, The Netherlands.}
\date{\today}
\begin{abstract}
The dynamics of drop impact on solid surfaces can be changed significantly by tuning the elasticity of the solid. 
Most prominently, the substrate deformation causes an increase in the splashing threshold as compared to impact onto perfectly rigid surfaces, and can thus lead to splash suppression. 
Here, we experimentally determine the splashing threshold for impact on thin membranes as a function of the tension in the membrane and its elastic properties. 
The drop dynamics is correlated to the membrane deformation, which is simultaneously measured using a laser profilometry technique. 
The experimental results enable us to adapt current models for splashing, showing quantitatively how substrate deformation alters the splashing threshold.
\end{abstract}
\maketitle
\section{INTRODUCTION}
Drop splashing onto an elastic membrane is a widely observed phenomenon. 
For example, it is encountered when rain drops impact onto leaves, where the impact can lead to damage~\cite{Gart2015}, or the splashing drop can play a role in the outbreak of foliar diseases~\cite{Gilet2015}. 
In industry, drop splashing onto an elastic substrate is encountered in e.g. pesticide delivery~\cite{Peirce2016}, (biological) inkjet printing~\cite{Tirella2011, vanDam2004}, and (cold) spray coating~\cite{Lupoi2010, Srikar2009}. 
In these applications, splashing decreases the deposition efficiency and may lead to widespread contamination, and is therefore an unwanted side effect.

Splashing of drops during impact is widely studied, as a function of the drop properties~\cite{Mundo1995, Xu2007, Rioboo2002}, the properties of the surroundings~\cite{Xu2005, Latka2012}, and the substrate properties~\cite{Tsai2010, Latka2012, Xu2007, Pepper2008, Howland2016}. 
The impact velocity~\cite{Mundo1995} or viscosity~\cite{Xu2007} of the drop are examples of drop properties that can be changed to control splashing.
The surrounding air influences drop splashing by its pressure~\cite{Xu2005, Latka2012}.
The substrate at which the drop impacts also influences drop splashing, for example by its structure~\cite{Tsai2010, Latka2012, Xu2007}.

To quantify the splashing threshold of a drop with diameter $D$, velocity $U$, density $\rho$, surface tension $\gamma$, and kinematic viscosity $\nu$, \citeauthor{Mundo1995}\cite{Mundo1995} derived the well-known criterion: We$^{1/2}$Re$^{1/4}> K$, where We = $\frac{\rho DU^2}{\gamma}$ is the Weber number, Re = $\frac{DU}{\nu}$ is the Reynolds number, and $K$ is the critical number to obtain splashing depending on for example substrate properties.
This splashing criterion can be related to the argument that the velocity at which the ejecta sheet moves away from the impact location has to be larger than the Taylor-Culick velocity to obtain splashing~\cite{Josserand2003}. 
For different kinds of substrates, typical values for $K$ are found: $K \approx$ 54 for a rigid surface~\cite{Josserand2016}, $K \approx$ 160 for a thin liquid film~\cite{Deegan2008} and $K \approx$ 90 for a deep liquid pool~\cite{Castillo2015}.
For oblique drop impact on a liquid pool, this splashing criterion is modified to account for the parallel impact velocity~\cite{Gielen2017}.
However, this splashing criterion by \citeauthor{Mundo1995} does not account for the role of the surrounding atmosphere~\cite{Xu2005, Latka2012}.

Recently, \citeauthor{Riboux2014}~\cite{Riboux2014} derived a new splash criterion that includes the effect of the surrounding gas and showed good agreement with the experimental results of Ref.~\cite{Xu2005}.
In this argument the liquid not only has to dewet the solid, but the upward velocity of the ejecta also has to be large enough to prevent touchdown of the growing rim, which leads to an adapted splashing threshold:
\begin{equation}
(\mu_gU/\gamma)\tilde{t}^{-1/2}_{e} > C, \label{eq:C}
\end{equation}
where $C$ is the critical number to obtain splashing based on e.g. surface properties, $\mu_g$ is the dynamic viscosity of the surrounding gas and $\tilde{t}_{e}=t_eU/D$ is the dimensionless time for sheet ejection.

On elastic or soft substrates, the splashing threshold is found to increase with decreasing substrate (visco)elasticity~\cite{Pepper2008, Howland2016}.
On a viscoelastic substrate~\cite{Howland2016} a drop needs up to 70 \% more kinetic energy to splash compared to a rigid surface, which is attributed to the substrate deformation during the early stages of impact.
A similar result is found on elastic membranes~\cite{Pepper2008}.
However quantitative data of the substrate deformation and supporting modeling are lacking. 
Moreover, the role of substrate elasticity has not been evaluated in light of the recent advances on the splashing threshold by Ref.~\cite{Riboux2014}.

Here, we experimentally study the splashing threshold on an elastic membrane. 
Details of the experimental setup for analyzing the drop impact and the membrane deformation are provided in Sec.~\ref{Exp} and a typical impact experiment is discussed in Sec.~\ref{Res1}. 
The splashing threshold is measured for different membrane elasticities in Sec.~\ref{Res2}.
Systematic measurements of the membrane deformation are presented in Sec.~\ref{Res3}.
In Sec.~\ref{Int} we interpret our observations and modify the splashing threshold model by Ref.~\cite{Riboux2014}, taking the elastic properties of the membrane into account.
The paper closes with a discussion in Sec.~\ref{Disc}.

\section{EXPERIMENTAL METHODS} \label{Exp}
\begin{figure}
\centering
\begin{subfigure}[b] {\textwidth}
\includegraphics[width=0.6\textwidth]{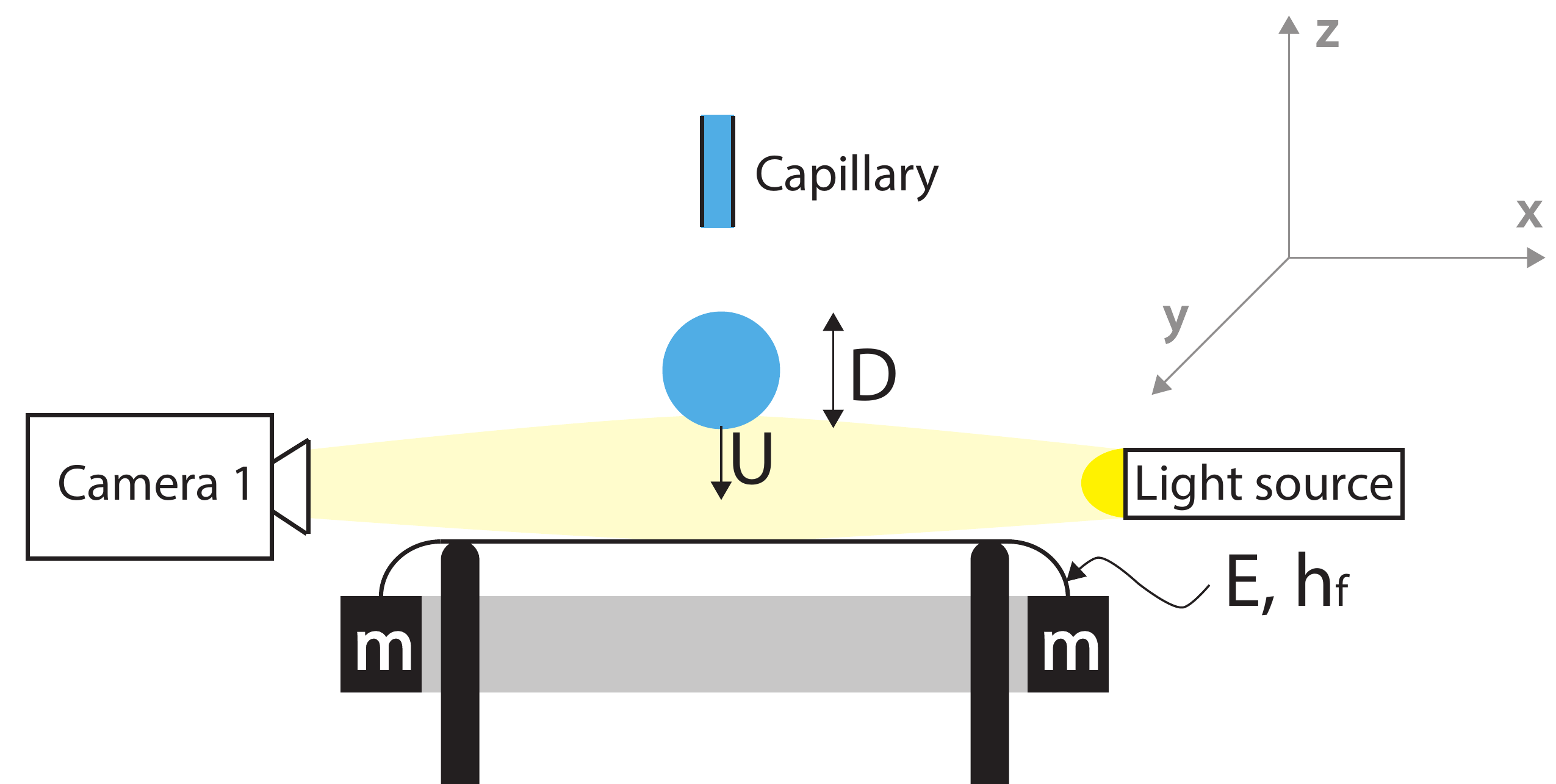}
\caption{}
\label{fig:impact}
\end{subfigure}\\
\begin{subfigure}[b] {\textwidth}
\includegraphics[width=0.6\textwidth]{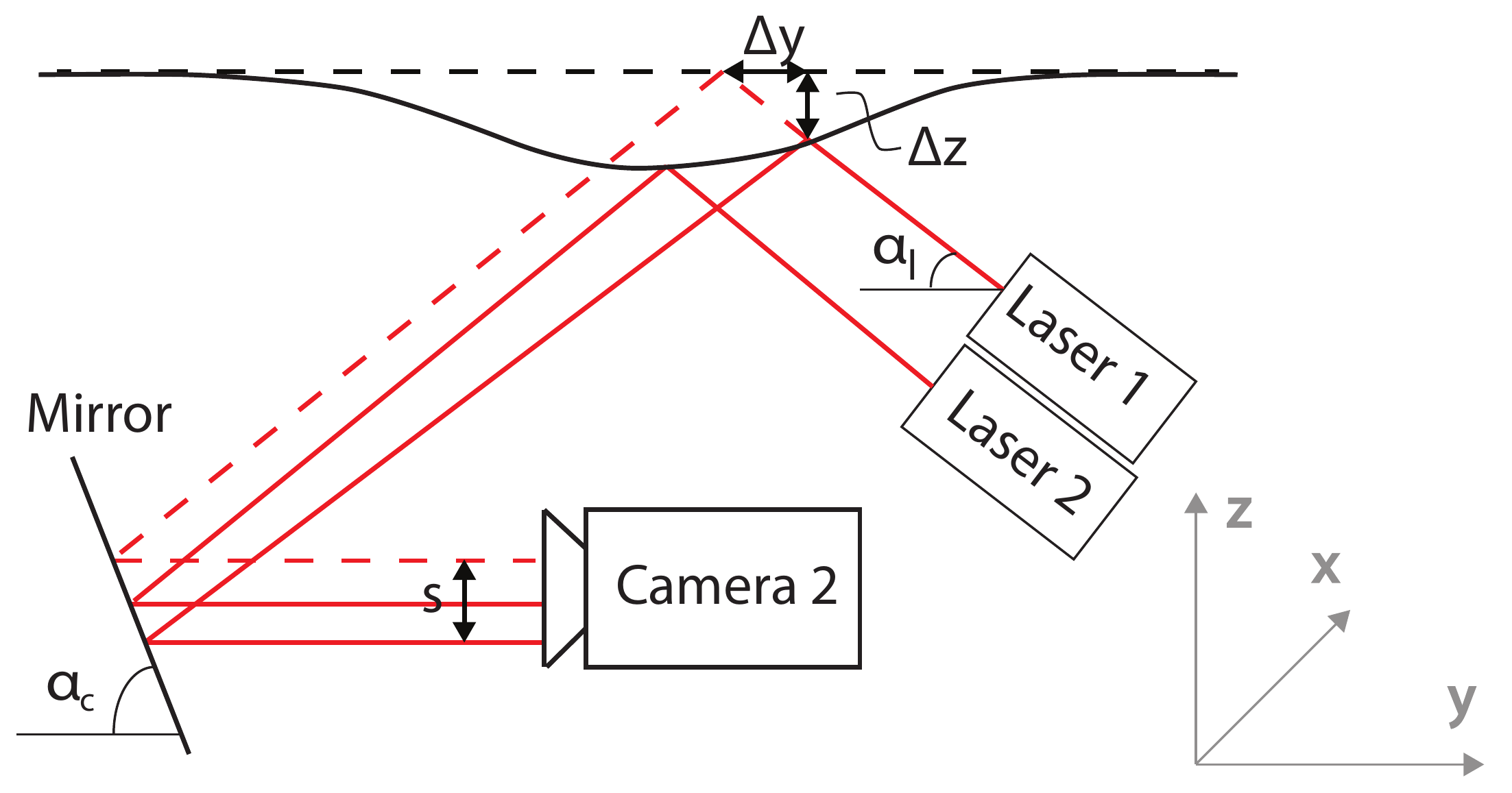} 
\caption{}
\label{fig:deformation}
\end{subfigure}
\caption{(a) Sketch of the drop impact setup. 
At the end of the capillary an ethanol drop with diameter $D$ is created. 
The drop falls down under influence of gravity, such that by changing the release height of the drop the impact velocity $U$ is controlled. 
The drop impacts onto an elastic membrane, where the added mass $M$ determines the variable tension (9 $\leq T \leq$ 46 N/m).
The membrane has thickness $h$ = 13 $\mu$m and Young's Modulus $E$ = 180 MPa. 
A side-view recording of the impact event is made using high speed shadowgraphy imaging. 
(b) To measure the membrane deformation two laser sheets are focused at the bottom of the membrane within a drop radius apart from each other. 
A second high speed camera focuses onto the membrane, via a mirror. 
When a drop impacts the membrane deforms and the recorded position of the laser sheets changes.
The displacement $s$ of the laser sheets on the camera is related to the actual membrane deformation $\Delta y$ and $\Delta z$ through geometrical relations, using laser angle $\alpha_\ell$ and camera angle $\alpha_c$~(Eq.~(\ref{eq:dxdy})).}
\label{fig:schematic}
\end{figure}

To study the influence of membrane elasticity on drop splashing during impact, we place an elastic membrane (plastic wrap, WRAPPITT) over a circular Teflon frame with diameter 10 cm, as shown in Fig.~\ref{fig:schematic}(a).
The setup is similar to the one described in Ref.~\cite{Pepper2008}, but with choice of different materials for frame and membrane. 
The thickness of the membrane $h$ = 13 $\pm$ 0.5 $\upmu$m, the density $\rho_f$ = 1.7$\cdot$10$^3$ kg/m$^3$, the Young's Modulus $E$ = 180 $\pm$ 37 MPa (calculated from measurements of the extension of the plastic wrap under the influence of added mass) and the roughness $R_a$ $\leq$ 50 nm (as measured with SensoSCAN, Sensofar metrology).
The frame is wetted to allow the membrane to slide over the frame during the experiments.

The membrane is placed under tension by attaching it to a mass (Fig.~\ref{fig:schematic}(a)). 
The mass is added to the membrane carefully, where care is taken to apply no additional tension. 
To vary the tension in the membrane the mass $M$ attached to it is varied between 292 and 1460 g. 
The tension is then given by $T=Mg/2\pi R_f$, with $g$ the gravitational constant and $R_f$ the radius of the frame and ranges from 9 $\pm$ 0.2 $\leq$ $T$ $\leq$ 46 $\pm$ 0.7 N/m.
To obtain a measure for the error in the applied additional tension, we  use multiple membranes for the experiments at some tensions ($T\approx$ 9, 27 and 46 N/m).
We notice that when putting the membrane under tension it shows a considerable amount of creep, which affects the reproducibility of our experiments. 
Therefore, we wait at least two hours after we place the membrane over the frame before performing the experiment.

An ethanol drop with a diameter of $D$ = 2.3 $\pm$ 0.4 mm is created by pumping ethanol at low flow rate through a capillary. 
Experiments take place under ambient conditions, such that the ethanol density $\rho$ = 789 kg/m$^3$ and the surface tension $\gamma$ = 0.02 N/m. 
By varying the release height of the drop, the impact velocity is controlled between 0.7 $\leq U \leq$ 5.1 m/s.
The outcome of the impact is recorded by side-view shadowgraphy imaging. 
The drop is illuminated (Sumita LS-M352A) and recordings are taken at 8,000 fps (Photron SA-X2) at a resolution of 23 $\mu$m per pixel.

To relate the drop dynamics to the membrane deformation, we simultaneously perform shadowgraphy and membrane deformation measurements.
The deformation is measured by laser profilometry, as decribed by Refs.~\cite{Pepper2008} and~\cite{Zhao2015}.
The setup for the profilometry measurements is shown schematically in Fig.~\ref{fig:schematic}(b).
We focus two laser sheets (Laser Macro Line Generator, Sch\"after+Kirchhoff, $\lambda$ = 660 $\pm$ 5 nm, sheet width $\sim$ 300 $\mu$m) from below at the bottom of the membrane such that the impacting drop is not blocking the view. 
The use of two laser sheets makes the measurement less sensitive to small variations in the exact impact locations, and provides an additional measurement. 
A second camera (Photron SA1.1) is focused via a mirror onto the bottom of the membrane recording the location of the laser sheets.
In order to increase the reflection of the membrane, the bottom of the membrane is coated with a white powder.
Because of this coating, the intensity of the reflections of the laser sheets increases and the quality of the recording improves, obtaining a resolution of 21 $\mu$m per pixel at a frame rate of 8,000 fps.

During drop impact the membrane will move down, resulting in a displacement of the laser sheets on the camera.
This displacement is related to the actual membrane deformations $\Delta y$ and $\Delta z$ via geometrical relations (Fig.~\ref{fig:schematic}(b)).
For small deformations, the deformation is linearly related to the displacement of the laser sheet $s$~\cite{Zhao2015},
\begin{subequations}
\begin{equation}
\Delta z = \frac{s}{2}\frac{\sin(\alpha_c+\alpha_\ell)}{\cos(\alpha_\ell)}
\end{equation}
\begin{equation}
\Delta y = \frac{s}{2}\frac{\sin(\alpha_c+\alpha_\ell)}{\sin(\alpha_\ell)},
\end{equation}
\label{eq:dxdy}%
\end{subequations}
\noindent{}where $\alpha_\ell$ is the angle of the laser and $\alpha_c$ is the angle of the camera, corresponding to the angle of the mirror.

To calculate $\Delta y$ and $\Delta z$ correctly, $\alpha_\ell$ and $\alpha_c$ are determined from separate  calibration measurements.
For $\alpha_\ell$, we track the direction of the laser sheets using our drop impact shadowgraphy setup as the light passes through the membrane. 
For $\alpha_c$ we remove the membrane from our frame and use a flat target for calibration.
For various target heights, we know $\Delta z$ from our shadowgraphy measurement and from our profilometry setup we know $\alpha_\ell$ and measure $s$, which allows us to calculate $\alpha_c$ from these combined experiments.

\begin{figure*}
\centering
\begin{subfigure}[t]{0.25\textwidth}
\centering
\includegraphics[width=\textwidth]{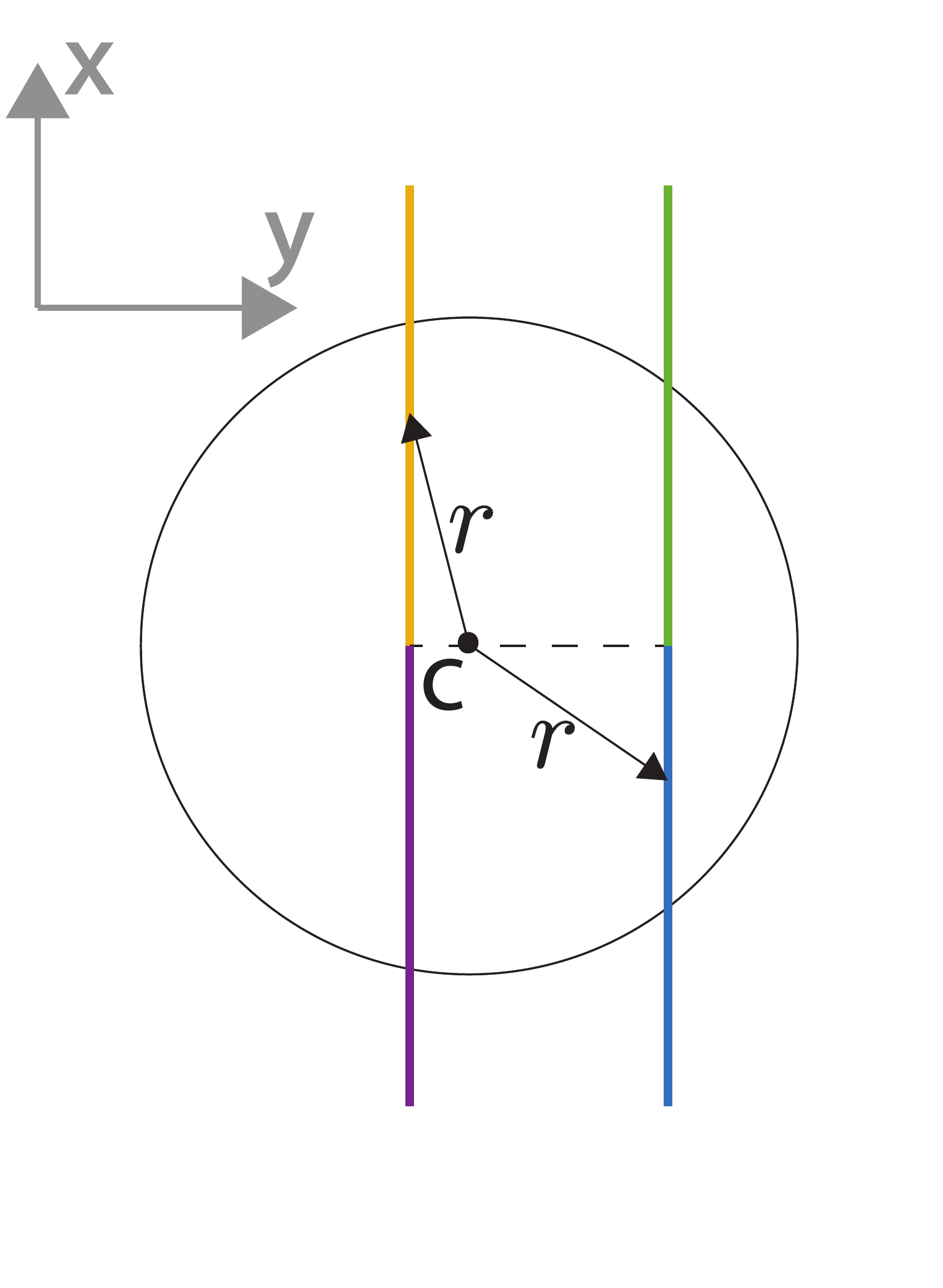}
\caption{}
\label{fig:LI}
\end{subfigure}
\begin{subfigure}[t]{0.5\textwidth}
\centering
\includegraphics[width=\textwidth]{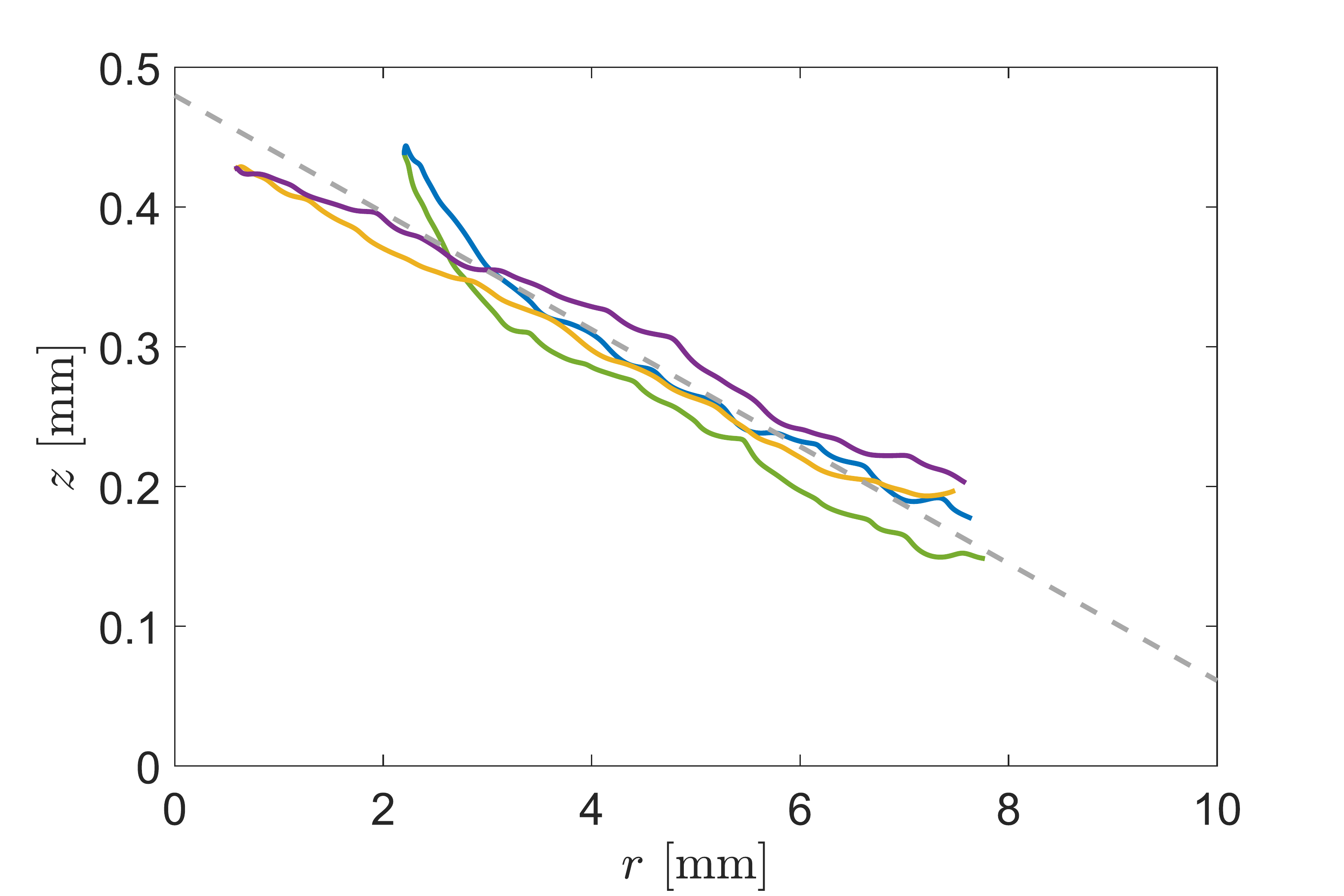} 
\caption{}
\label{fig:zr}
\end{subfigure}
\caption{(a) Schematic representation of the drop impact center $c$ with respect to the location of both laser sheets.
By combining the shadowgraphy and profilometry recordings, the location of the impact center $c$ is determined. 
From here the radial distance to both laser sheets is calculated.
Each laser sheet therefore consists of two branches starting at the same radial position.
(b) Plot showing the membrane profile at maximum deformation for We = 1123 and $T/Eh$ = 0.012 (see Fig.~\ref{fig:Series}). 
The colors match the branches from Fig.~\ref{fig:zrall}(a).
With increasing radius from the impact center the deformation decreases.
A linear fit (dashed line) is made to the laser sheets to find the maximum deformation at the center of the impacting drop ($r=0$).}
\label{fig:zrall}
\end{figure*}

For every impacting drop we detect the profile of the laser sheets.
The maximum deformation along the laser sheet gives the impact center of the drop in $x$-direction (see Fig.~\ref{fig:zrall}(a)), while the impact center of the drop in $y$-direction is extracted from the shadowgraphy measurements.
Once we know the exact impact position of the drop, we relate every detected point of the laser sheet to a radial distance from the impact center, as shown in Fig.~\ref{fig:zrall}(a).
Combining this radial distance and the measured deformation, we construct the entire profile of the deformed membrane.
An example is shown in Fig.~\ref{fig:zrall}(b).
The maximum deformation is not captured by the laser sheets, therefore a linear fit through the four branches is made to determine the maximum deformation at $r = 0$.
For each frame, the average value at $r = 0$ of those fits gives an estimate of the central deformation $\delta(t)$ of a single experiment.
We compared different fitting functions, and found the linear fit the most appropriate and robust, even though it lacks the condition of vanishing slope at $r=0$ and therefore slightly overestimates $\delta$. 

In the experiments we can vary the drop diameter $D$, impact velocity $U$ and mass $M$ (resulting in a change in membrane tension $T$) and measure the outcome of the impact and the deformation of the membrane $z$ simultaneously.
In dimensionless form, the control parameters are the Weber number We$=\rho DU^2/\gamma$ and the scaled tension $T/Eh$.
Typically, $T/Eh\ll1$ in our experiments.
Our goal is to identify the critical Weber number for splashing as a function of the dimensionless tension, and to correlate this to the membrane's deformability.
\section{RESULTS} \label{Res}
We first explain the general outcome of a drop impact event, emphasizing the information we get from the simultaneous shadowgraphy and profilometry measurements in Sec.\ref{Res1}.
In Sec.~\ref{Res2} the splashing threshold obtained from shadowgraphy measurements and its dependence on impact velocity and membrane tension are presented.
Finally, in Sec.~\ref{Res3} we show how the membrane deformation varies with impact velocity and membrane tension.
\subsection{A typical drop impact experiment} \label{Res1}
\begin{figure*}
\centering
\includegraphics[width=\textwidth]{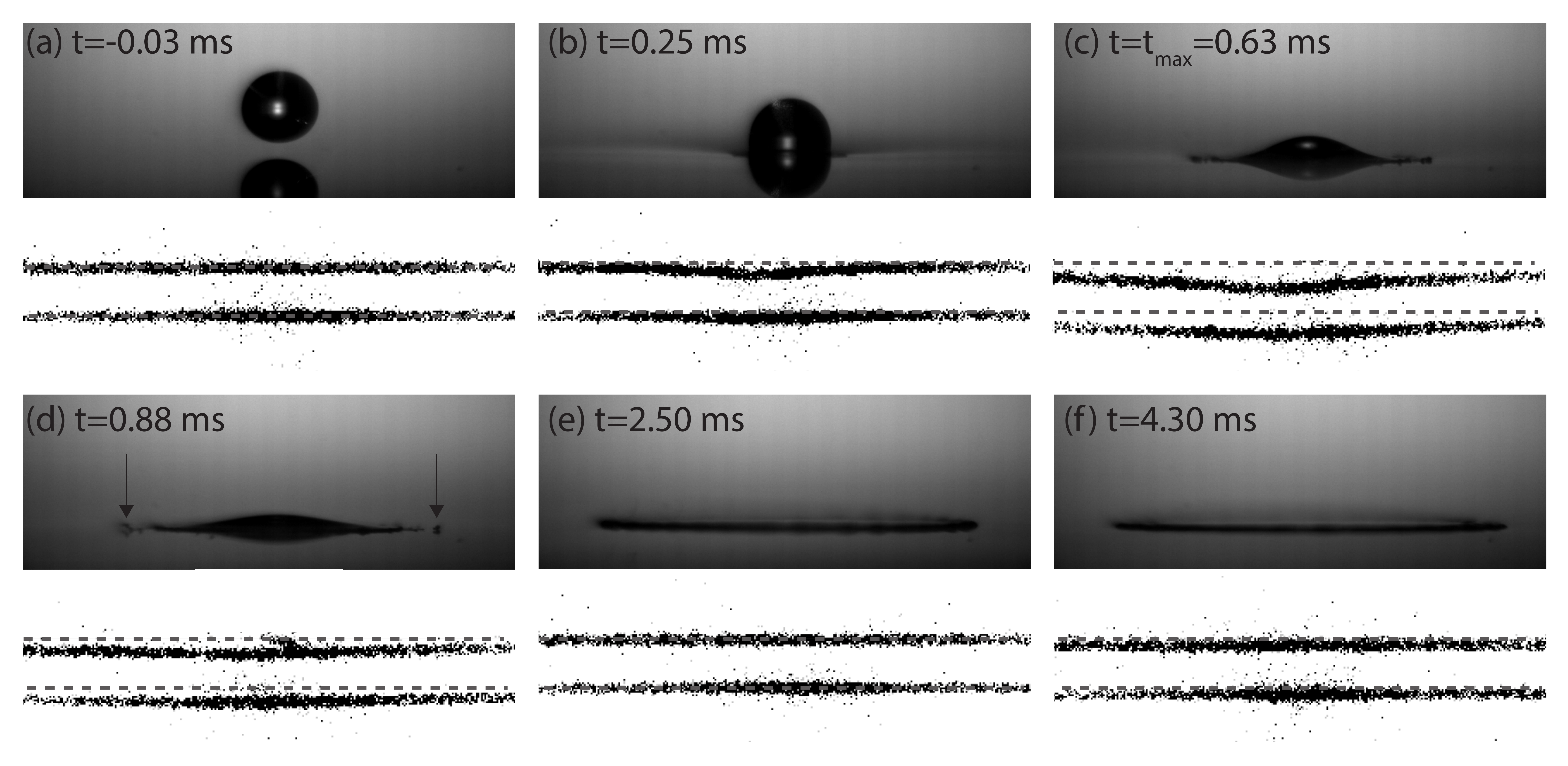} 
\caption{Time series of a drop impacting onto the membrane. 
The dimensionless tension in the membrane is $T/Eh$ = 0.012 and the impact Weber number We = 1123. 
Time t=0 marks the moment the drop makes first contact with the membrane. 
Each snapshot consists of two images: a side-view image where the drop is visible and an image of the laser sheet reflection on the membrane to show the deformation of the membrane. 
For clarity, the dashed lines indicate the original position of the laser sheet. 
(a) $t$ = -0.03 ms: Just before impact, when the membrane is undeformed. 
(b) $t$ = 0.25 ms: Right after the impact both the drop and the membrane start to deform. 
(c) $t$ = 0.63 ms: The drop spreads over the membrane, while the membrane reaches its maximum deformation. 
The corresponding membrane deformation is shown in Fig.~\ref{fig:zrall}(b).
(d) $t$ = 0.88 ms: Drop splashing; the formation of secondary droplets is observed, indicated by the arrows. 
(e) $t$ = 2.50 ms: The membrane has returned to its original position, while the drop is still spreading over the membrane. 
(f) $t$ = 4.30 ms: The drop reaches its maximum spreading, while the membrane oscillates around its equilibrium position.}
\label{fig:Series}
\end{figure*}
Figure~\ref{fig:Series} shows the time series of a typical experiment with We = 1123 and dimensionless tension $T/Eh$ = 0.012. 
The top row images show the drop during impact while the bottom row images show the measured deflection of the laser sheets at the same time.
Before impact (Fig.~\ref{fig:Series}(a)) the membrane is undeformed, while in Fig.~\ref{fig:Series}(b) the impact has started and both drop and membrane are deforming. 
At $t = t_{max}$ = 0.63 ms (Fig.~\ref{fig:Series}(c)), the maximum deformation of the membrane is reached, while the drop is still spreading.
Shortly after the moment of maximum deformation, secondary droplets break off from the initial drop and a splash is visible (Fig.~\ref{fig:Series}(d)). 
After $t$ = 2.50 ms (Fig.~\ref{fig:Series}(e)), the membrane has released all of its stored energy and has returned to its original position.
The drop, however, is still spreading further over the membrane until it reaches its maximum spreading diameter at $t$ = 4.30 ms (Fig.~\ref{fig:Series}(f)).
During this spreading, the membrane oscillates around its equilibrium position until these oscillations are damped out (images not shown here).
An important observation, which applies to all splashing events studied, is that the splash occurs after the membrane has reached its maximum deformation.
This observation is consistent with the result of Ref.~\cite{Pepper2008}.
\subsection{Splashing threshold} \label{Res2}
\begin{figure*}
\centering
\begin{subfigure}[t]{0.49\textwidth}
\centering
\includegraphics[width=\textwidth]{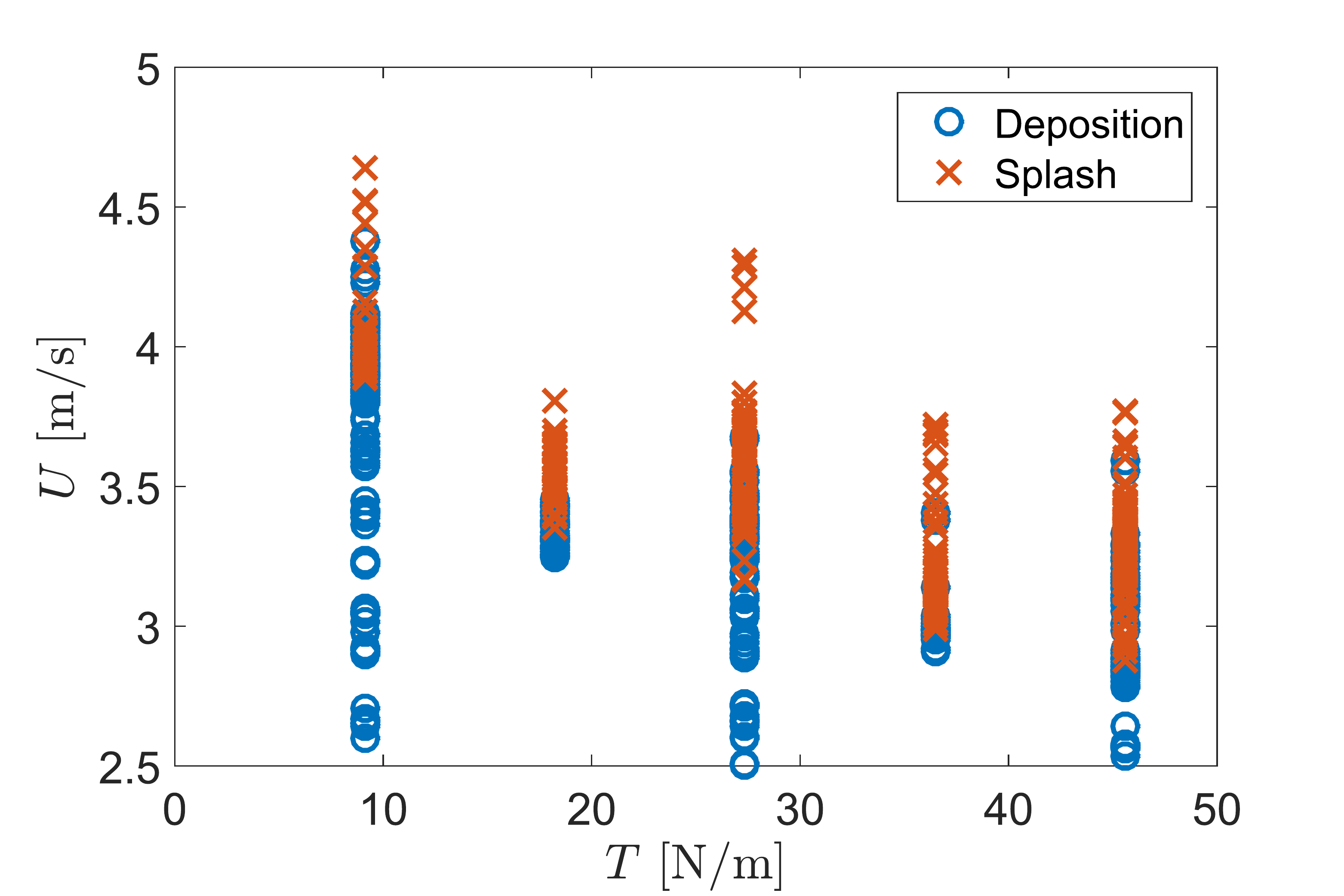}
\caption{}
\label{fig:Usigma}
\end{subfigure}
\begin{subfigure}[t]{0.49\textwidth}
\centering
\includegraphics[width=\textwidth]{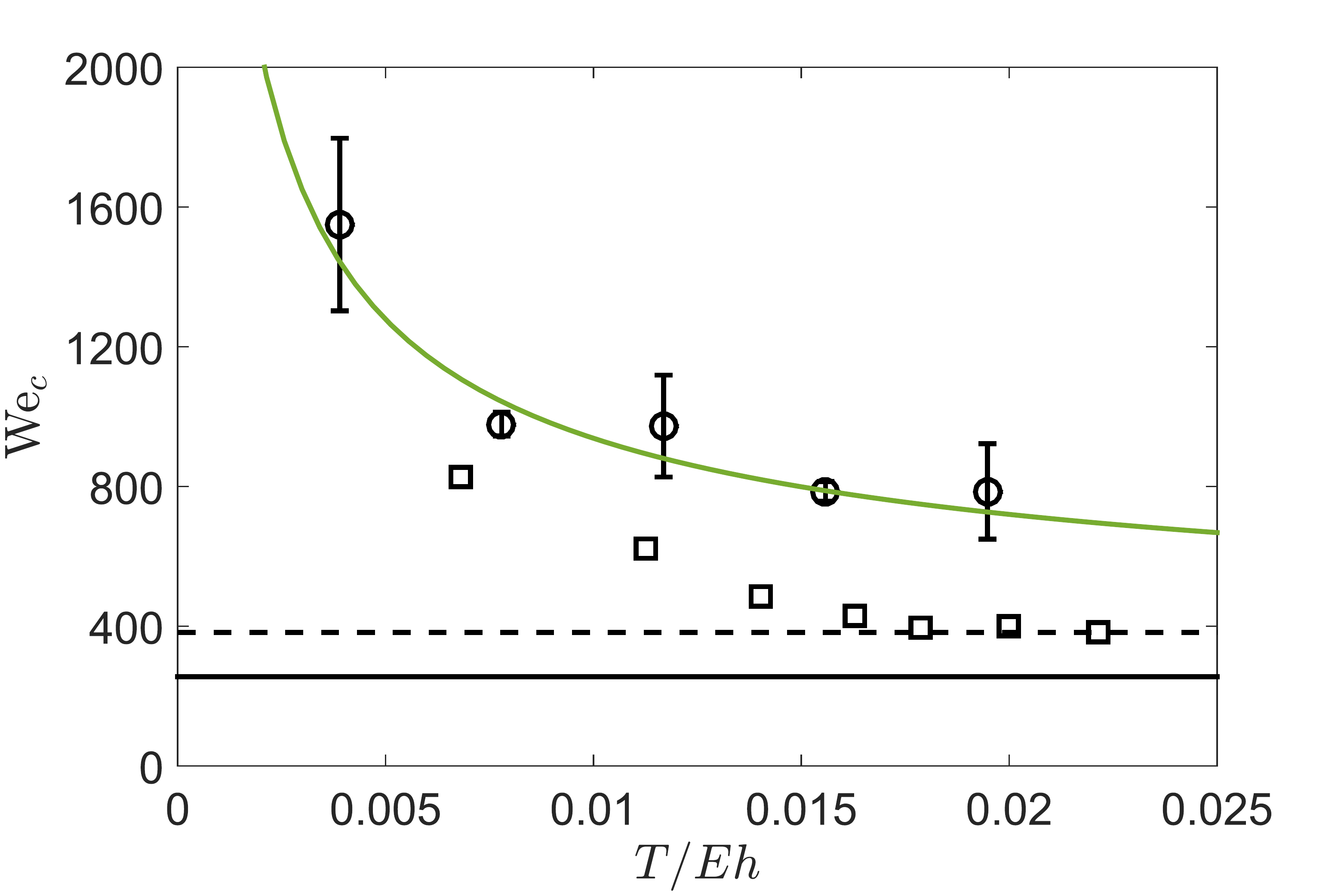} 
\caption{}
\label{fig:WecsigmaE}
\end{subfigure}
\caption{(a) Plot of all individual measurements done for different tensions in the membrane $T$ and varying impact velocity $U$. 
The blue circles indicate deposition experiments, while the red crosses indicate splashing experiments. 
For each tension we find a zone where deposition and splashing events overlap, from which we determine the splash threshold velocity $U_c$.
(b) Non-dimensional plot of the critical Weber number We$_c = \rho D U_c^2/\gamma$ for splashing as function of the non-dimensional tension in the membrane $T/Eh$. 
Circles represent measurements of this study, while squares represent measurements by Ref.~\cite{Pepper2008}.
The error bars indicate one standard deviation. 
The solid black line is the splashing threshold on a rigid substrate with the same surface properties (standard deviation negligibly small), the black dashed line indicates the rigid splashing threshold found by Ref.~\cite{Pepper2008}.
For decreasing $T/Eh$, We$_c$ increases.
The solid green line shows the splashing threshold according to Eq.~(\ref{eq:Cmem}) with $c_t = 0.07$ for our measurements.}
\label{fig:splashthres}
\end{figure*}

We now study the impact of a drop onto an elastic membrane by systematically varying the impact velocity and the membrane tension. 
Two different outcomes of an impact event are observed: 
(1)~Deposition: After impact the drop remains intact on the membrane.
(2)~Splashing: After impact small secondary droplets break off from the rim of the original drop.
In case only one or two droplets are ejected on the same side, this is still referred to as deposition.
Figure~\ref{fig:splashthres}(a) shows all experimental data for different impact velocities and tensions.
For low impact velocities, the impacting drop deposits onto the membrane, while for higher impact velocities the drop splashes during impact. 
With decreasing tension the impact velocity at which splashing is first observed increases.
For each tension there is a small region where both deposition and splashing are observed, which is called the transition region.
From this transition region the splashing threshold and its standard deviation are derived, which will be used as an error bar for the determination of the splashing threshold. 
To this end, the impact velocities are grouped in bins with a range of 0.25 m/s and the percentage of drops that splashes during impact is determined. 
A cumulative distribution function fit is made through all these data points to find the expected threshold value and its standard deviation.
To test the reproducibility of our results, we repeated the same experiments with a new membrane for $T/Eh$ = 0.0039, 0.012, and 0.020.
The increase in standard deviation of the splashing threshold for these experiments shows the sensitivity of the experiments to the manual fixation of the mass to the membrane.

The threshold for splashing is presented in non-dimensional form in Fig.~\ref{fig:splashthres}(b), where we show the critical Weber number We$_c = \rho DU_c^2/\gamma$ as function of the dimensionless tension $T/Eh$ (black circles).
As a reference case, we also performed experiments onto a membrane carefully placed on top of a glass plate, where we avoid air bubble entrainment between the glass plate and the membrane, finding a splashing threshold of We~=~255 (solid line). 
While We$_c$ for splashing decreases with increasing dimensionless tension, the rigid threshold is never reached, which shows that even at the highest tension impact energy is used to deform the membrane. 
In Ref.~\cite{Pepper2008} a similar range of tensions was studied and the results are reproduced here for comparison (black squares). 
The authors also found a decrease in splashing threshold, however, in contrast to our results, there the rigid limit (We~=~382) was reached already for the highest tensions.
Despite the higher rigid splashing threshold, the typical values reported for We$_c$ on the membrane are lower compared to our observations.
\subsection{Membrane deformation} \label{Res3}
To explain the splash suppression mechanism, we now turn to the membrane deformation.
From a simultaneous measurement of the impact event and membrane deformation, we can relate the deformation dynamics to the impact velocity of the drop.
In Fig.~\ref{fig:deltat}(a) we plot the central deformation (at $r$ = 0) as function of time.
At $t=t_{max}$ the maximum deformation $\delta_{max}$ is reached. 
Before $t=t_{max}$, the membrane deforms and stores part of the kinetic energy of the impacting drop as elastic energy.
For $t>t_{max}$, the membrane returns to its original position and (part of) the stored elastic energy is released.
At later times, the membrane oscillates and subsequently relaxes to its final position.
\begin{figure}
\centering
\begin{subfigure}[t]{0.49\textwidth}
\centering
\includegraphics[width=\textwidth]{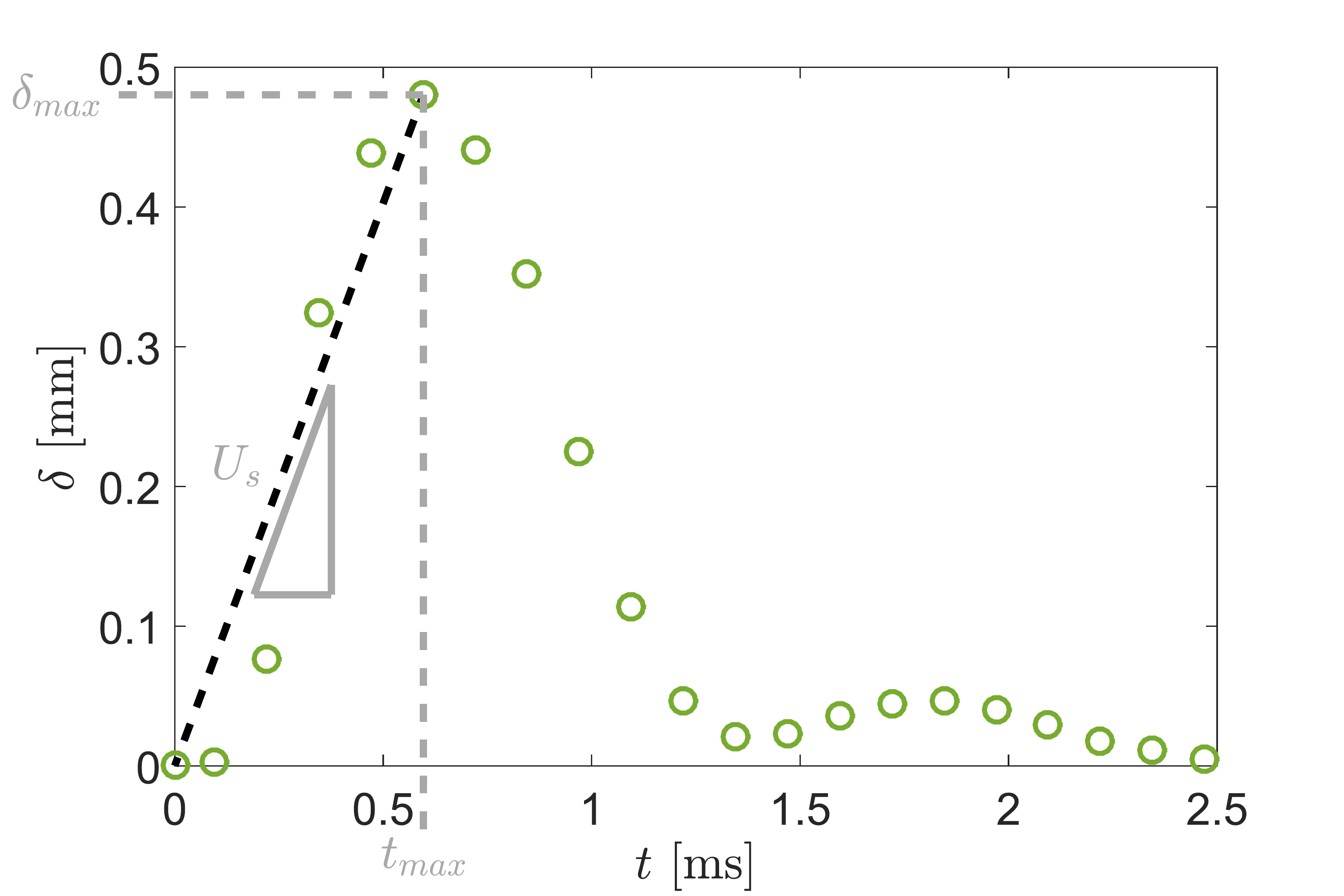}
\caption{}
\label{fig:deltat}
\end{subfigure}
\begin{subfigure}[t]{0.49\textwidth}
\centering
\includegraphics[width=\textwidth]{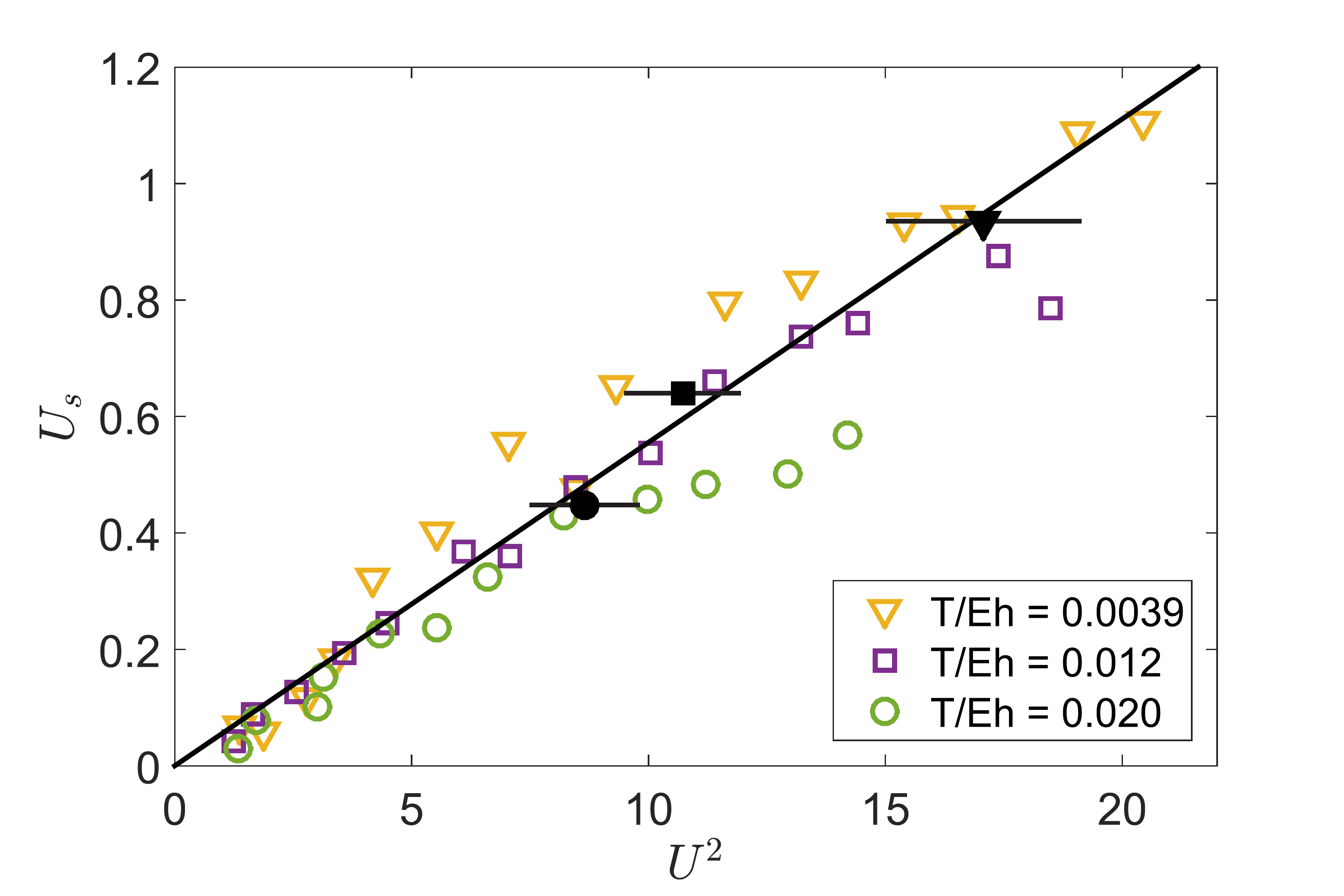} 
\caption{}
\label{fig:UsWe}
\end{subfigure}
\caption{(a) Plot of the central deformation of the membrane $\delta$ as function of time $t$ for the same experiment as shown in Fig.~\ref{fig:Series} (We = 1123 and $T/Eh$ = 0.17). 
Time $t=0$ corresponds to time of impact. 
A linear fit from zero to maximum deformation $\delta_{max}$ gives a measure for the velocity at which the membrane deforms (here $U_s \approx 0.77$ m/s) . 
Before $t = t_{max} = 0.63$ ms, kinetic energy of the impacting drop is stored as elastic energy in the membrane. 
For $t > t_{max}$, this elastic energy is released and the membrane returns to its original position while oscillating.
(b) Plot of the membrane velocity $U_s$ as a function of $U^2$ for three different tensions.
We have taken bins of $U$ = 0.25 m/s to group experiments.
$U_s$ is linearly related to $U^2$ until the splashing threshold is reached.
Black symbols indicate the critical velocity for splashing $U_c$ at different tensions with corresponding error bars.
The solid line shows the empirical fit given by Eq.~(\ref{eq:usu}) to the data with $c_U$ = 18.}
\label{fig:deltat}
\end{figure}•

From Fig.~\ref{fig:deltat}(a) one can extract the velocity $U_s$ at which the membrane initially deforms from a linear fit.
Figure~\ref{fig:deltat}(b) shows $U_s$ as function of $U^2$.
One observes that the membrane velocity increases linearly with $U^2$.
For larger impact velocities $U$ a deviation from this trend is observed, which starts at lower $U$ for higher tension.
The start of this deviation turns out to coincide with the splashing threshold $U_c$, marked as the black symbols in Fig.~\ref{fig:deltat}(b).
Apparently, from this impact velocity on the membrane can no longer store enough impact energy to prevent the drop from splashing.
The lower the tension, the higher $U_c$, and therefore the higher the membrane velocity where the deviation starts.
For $U<U_c$, the membrane velocity collapses onto a single linear curve (black solid line in Fig.~\ref{fig:deltat}(b)) for all tensions.
To describe the data up to the splashing threshold we therefore propose the following empirical law
\begin{equation}
U_s=c_U\frac{U^2}{U^*}, \label{eq:usu}
\end{equation}
where $U^*=\sqrt{E/\rho_f}$ is the characteristic wave velocity of the membrane and $c_U$ = 18 gives a fit to our data.

\begin{figure*}
\centering
\begin{subfigure}[t]{0.49\textwidth}
\includegraphics[width=\textwidth]{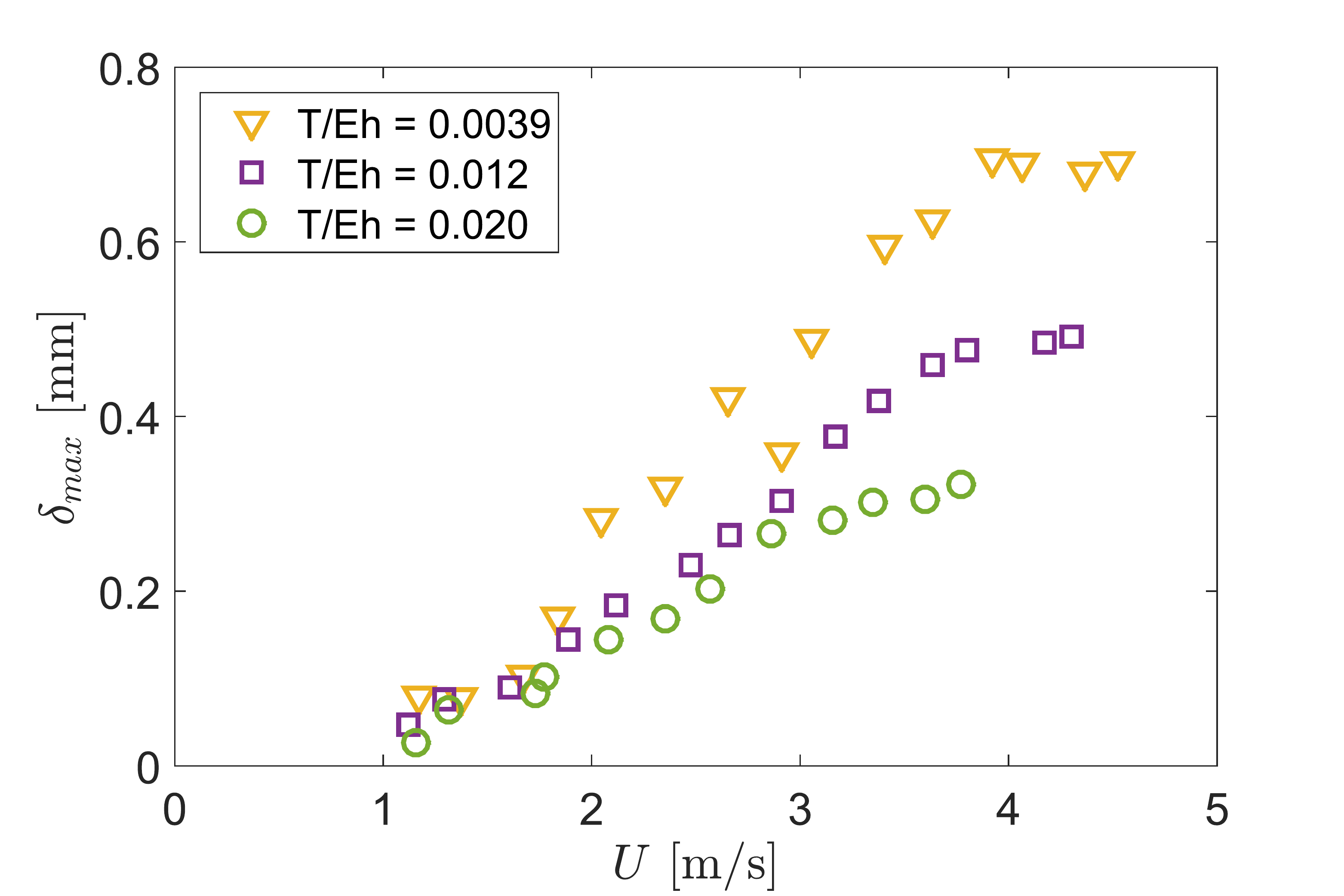} 
\caption{}
\label{fig:deltaWe}
\end{subfigure}
\begin{subfigure}[t]{0.49\textwidth}
\includegraphics[width=\textwidth]{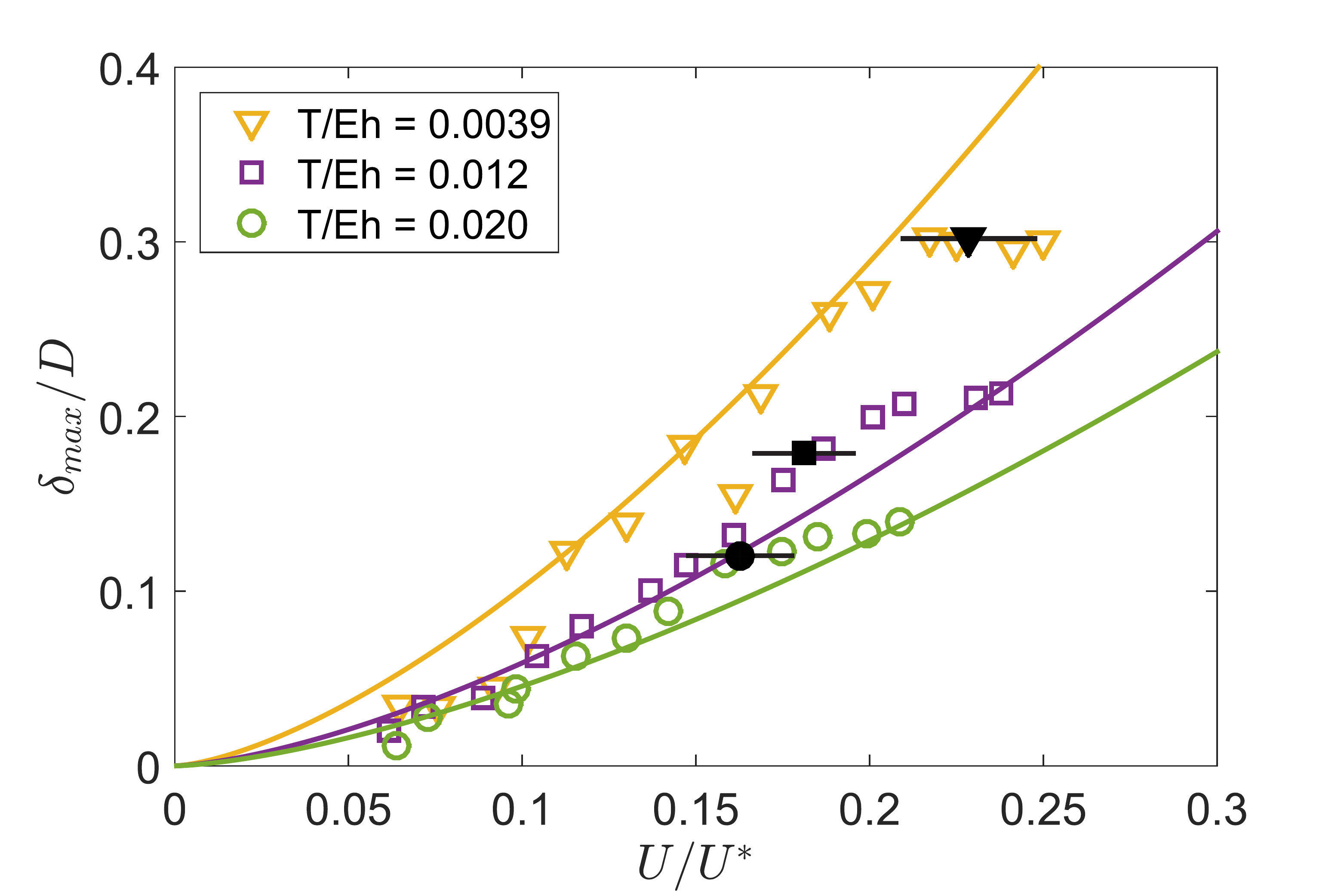} 
\caption{}
\label{fig:deltanon}
\end{subfigure}
\caption{(a) Maximum deformation $\delta_{max}$ as function of the impact velocity $U$.
Each open symbol corresponds to a set of experiments averaged over a small range of $U$ (0.25 m/s).
For increasing impact velocity the maximum deformation increases. 
For lower membrane tension the deformation is larger at the same impact velocity.
(b) The same data as shown in (a) plotted in dimensionless form.
The solid black symbols indicate the maximum deformation measured at the splashing threshold.
The increase of deformation with impact velocity is captured by Eq.~(\ref{eq:expand}) with $c_{\delta}$ = 0.4.
After the splashing threshold a deviation is observed between model and experimental data.}
\label{fig:delta}
\end{figure*}

From the membrane deformation measurements as shown in Fig.~\ref{fig:deltat}(a) one can also extract the maximum deformation of the membrane.
In Fig.~\ref{fig:delta}(a) we show that the maximum deformation increases with impact velocity, as was to be expected.
In addition, the maximum deformation is larger when the tension is smaller.
•
\section{INTERPRETATION} \label{Int}
We now interpret and quantify the increase in splashing threshold with decreasing tension observed in Fig.~\ref{fig:splashthres}(b).
To this end, the model for the splashing threshold on a rigid surface as given by Ref.~\cite{Riboux2014} is modified to account for the membrane elasticity.
The modified threshold will turn out to involve both the membrane deformation velocity $U_s$ (Sec.~\ref{Res3}) and the maximum deformation $\delta_{max}$, which is therefore discussed in more detail in Sec.~\ref{dmax}.
In Sec.~\ref{st} we then use $U_s$ and $\delta_{max}$ to arrive at a splashing criterion for impact on an elastic membrane.

\subsection{Maximum membrane deformation} \label{dmax}
In Fig.~\ref{fig:delta}(b) we plot the dimensionless maximum deformation $\delta_{max}/D$ versus the impact velocity nondimensionalized by the characteristic membrane velocity $U/U^*$.
To understand the dependence of $\delta_{max}$ on $U$ and $T$, we propose a model for the maximum deformation based on momentum conservation in two consecutive steps:
$\textcircled{A}$ The momentum of the drop before impact is equated to the momentum of the drop and the membrane at the moment of the inelastic collision.
$\textcircled{B}$ Subsequently, at maximum deformation the drop has stopped moving downwards and the membrane has provided an impulse $\sim \sigma h\delta_{max}t_{max}$, with $\sigma=E\epsilon+T/h$ the total stress in the membrane.
In terms of a scaling analysis, the momentum balance then reads
\begin{equation}
\rho D^3U \overset{\textcircled{A}}{\sim} (\rho D^3 + \rho_fh\ell^2)U_s \overset{\textcircled{B}}{\sim} \left(E\epsilon+\frac{T}{h}\right)h\delta_{max}t_{max},
\label{eq:balance}
\end{equation}
where $\ell$ is the horizontal length scale over which the membrane is deformed and $\epsilon \sim (\delta_{max}/\ell)^2$ is the strain in the membrane. 
From \textcircled{A} we obtain an expression for $\ell$, making use of our experimental observation $U/U_s\gg 1$,
\begin{equation}
\ell^2\sim\frac{\rho D^3}{\rho_fh}\frac{U}{U_s}. \label{eq:l}
\end{equation}• 
For the parameters in our experiments this predicts $\ell\sim10^{-2}$ m, which is consistent with the experimental observations (Fig.~\ref{fig:zrall}(b)).

Eliminating $\ell$ through Eq.~(\ref{eq:l}) and using that $t_{max}\sim\delta_{max}/U_s$ in Eq.~(\ref{eq:balance})\textcircled{B} we obtain a quadratic relation for $\delta_{max}$:
\begin{equation}
\left(\frac{\delta_{max}}{D}\right)^2 = 
\frac{1}{2}c_\delta^2\frac{\rho D}{\rho_fh}\frac{T}{Eh}\frac{U}{U_s} \left[-1
+\sqrt{1 + 4\frac{Eh^2\rho_f}{T^2}U_s^2}\right].
\label{eq:delta}
\end{equation}
The use of an equality sign in Eq.~(\ref{eq:delta}) requires a definition of the proportionality constant $c_\delta$, which will be determined from experiments.
Noting that $4\tfrac{Eh^2\rho_f}{T^2}U_s^2 \ll 1$ allows us to reduce Eq.~(\ref{eq:delta}) to
\begin{equation}
\left(\frac{\delta_{max}}{D}\right) = 
c_\delta\left[\frac{\rho DUU_s}{T}\right]^{1/2},
\label{eq:expand}
\end{equation}•
where $U_s$ follows from Eq.~(\ref{eq:usu}).
A fit of Eq.~(\ref{eq:expand}) to the data in Fig.~\ref{fig:delta}(b) gives $c_{\delta}$ = 0.4.

Figure~\ref{fig:delta}(b) confirms that Eq.~(\ref{eq:expand}) captures the dependence of $\delta_{max}$ on both $U$ and $T$ up to $U_c$, where the membrane velocity starts to deviate from Eq.~(\ref{eq:usu}).

\subsection{Splashing threshold}\label{st}
The next step is to adapt the splashing criterion by \citeauthor{Riboux2014} given in Eq.~(\ref{eq:C}) to account for the membrane elasticity.
To this end, we use the empirical relation of the membrane velocity Eq.~(\ref{eq:usu}) and the model for the membrane deformation Eq.~(\ref{eq:expand}).

For completeness, we first briefly recap the model for splashing on a rigid substrate.
To obtain splashing, two criteria have to be fulfilled simultaneously~\cite{Riboux2014}:
(i) The ejecta sheet has to dewet the solid such that the ejecta sheet moves upwards, and
(ii) the vertical velocity at the tip of the ejecta sheet has to be large enough to prevent touchdown of the growing rim on the substrate.
Once both conditions are fulfilled, the splashing criterion is given by Eq.~(\ref{eq:C}), where the time scale for sheet ejection at the splashing threshold $t_{e,c}$ is found from~\cite{Riboux2014}
\begin{equation}
k_1\textrm{Re}_c^{-1}\tilde{t}_{e,c}^{-1/2} + \textrm{We}_c^{-1} = k^2\tilde{t}_{e,c}^{3/2},
\label{eq:te}
\end{equation}•
with $\tilde{t}_{e,c}=t_{e,c}U_c/D$ and $k_1=\sqrt{3}/2$ and $k=1.1$ taken from Ref.~\cite{Riboux2014}.
Our measurements on a rigid substrate coated by the elastic membrane give a critical impact velocity $U_c\approx$ 1.8 m/s.
Applying the model of Eqs.~(\ref{eq:C}) and~(\ref{eq:te}) to our measurements of a membrane coated glass slide, we obtain $\tilde{t}_{e,c}\approx$ 0.027 ($t_{e,c} \approx$ 35 $\mu$s), and find for the critical splashing number $C \approx$ 0.0093.

On an elastic membrane criteria (i) and (ii) still need to be fulfilled to obtain splashing.
However, as the impact energy is firstly absorbed by the membrane and released only at later times $t>t_{max}$, we hypothesize it is now the time scale of the membrane deformation that controls the sheet ejection.
We therefore suggest a modified splashing criterion by simply adding $t_e$ and $t_m$ in Eq.~(\ref{eq:C}), such that the splash criterion now reads
\begin{equation}
\frac{\mu_gU_c}{\gamma}(\tilde{t_e}+c_t \tilde{t}_{max,c})^{-1/2} = C,
\label{eq:Cmem}
\end{equation}•
where $c_t$ is a fitting parameter.
The critical dimensionless time of membrane deformation is given by $\tilde{t}_{max,c} = \tfrac{\delta_{max}}{D}\tfrac{U_c}{U_s}$, where the membrane velocity is given by Eq.~(\ref{eq:usu}) and the maximum substrate deformation by Eq.~(\ref{eq:expand}) with the prefactors as determined from the independent deformation measurements.
In the limit where $T\rightarrow\infty$ ($\tilde{t}_{max}\rightarrow 0$) one recovers the splashing criterion on a rigid substrate (Eq.~(\ref{eq:C})).

In Fig.~\ref{fig:splashthres}(b), Eq.~(\ref{eq:Cmem}) is compared to the experimental data, as shown by the solid green line.
In Eq.~(\ref{eq:Cmem}) $C$ = 0.0093 as obtained for a rigid substrate is used, since the only substrate property that is varied in the experiment is the tension.
We then solve Eq.~(\ref{eq:Cmem}) numerically, using $\mu_g$ = $1.7\cdot10^{-5}$ kg/m$\cdot$s and find good agreement with the experimentally determined splashing threshold for $c_t$ = 0.07.

\section{Discussion and Conclusion} \label{Disc}
The splashing threshold for drop impact onto an elastic membrane is determined by the simultaneous measurement of the impact behavior by side-view imaging and the membrane deformation by laser profilometry. 
The observed increase in splashing threshold with decreasing tension is interpreted by modifying the splashing criterion by Ref.~\cite{Riboux2014} to account for the membrane deformation.

From the profilometry measurements we extracted two key quantities that determine the splashing threshold: 
the membrane deformation velocity and the maximum membrane deformation.
We observed a quadratic scaling of the membrane velocity with the impact velocity independent of membrane tension, up to the splashing threshold.
For larger impact velocities the membrane can no longer deform fast enough and the drop splashes.
The maximum membrane deformation could accurately be described by a simple scaling argument based on momentum conservation, again up to the splashing threshold.
The increased splashing threshold is then explained by a delayed sheet ejection time caused by the membrane deformation dynamics, and showed good agreement with experimental data.
Unfortunately we could not test the splashing criterion against the data by Ref.~\cite{Pepper2008}, since the reported measurements do not provide enough data on the membrane deformation.

So far, we validated the modified splashing criterion by measuring the splashing threshold as a function of membrane tension. 
A more direct validation would require measurements of the sheet ejection time as a function of the membrane tension, which unfortunately was obscured from view by the membrane deformation. 
On visco-elastic gels similar problems have been reported~\cite{Howland2016}.
A detailed numerical model could be used to overcome these experimental limitations.
Moreover, such model could provide a detailed description of the membrane dynamics and explain the observed deviation from the quadratic scaling in the membrane velocity.
However, our basic modification of \citeauthor{Riboux2014}~\cite{Riboux2014} based on simple scaling arguments is already capable of predicting the observed trends.

\begin{acknowledgments}
This work is part of an Industrial Partnership Programme of the Netherlands Organization for Scientific Research (NWO). 
This research programme is co-financed by ASML.
We acknowledge R. de Jong and S.-C. Zhao for help in setting up the laser profilometry experiments and analysis.
\end{acknowledgments}
\bibliography{membrane_bib}

\end{document}